\documentstyle[12pt]{article}
 
\newcommand{\be}{\small\begin{equation}}
\newcommand{\ee}{\end{equation}\normalsize\vspace*{-0.1ex}}
\newcommand{\bea}{\small\begin{eqnarray}}

\newcommand{\eea}{\end{eqnarray}\normalsize\vspace*{-0.1ex}}
\newcommand{\bdm}{\small\begin{displaymath}}
\newcommand{\edm}{\end{displaymath}\normalsize\vspace*{-0.1ex}}
\newcommand{\beas}{\small\begin{eqnarray*}}

\newcommand{\eeas}{\end{eqnarray*}\normalsize\vspace*{-0.1ex}}
\newcommand{\n}{\noindent}
\newcommand{\MS}{\overline{\rm MS}}
\newcommand{\pole}{{\rm pole}} 
 
\begin{document}
 
 
\thispagestyle{empty}
\renewcommand{\thefootnote}{\fnsymbol{footnote}}
 
\setcounter{page}{0}
\begin{flushright} UM-TH-94-31\\
August 1994\\
hep-ph/9408380  \end{flushright}
 
\begin{center}
\vspace*{2.5cm}
{\Large\bf More on ambiguities in the pole mass}\\
\vspace{1.8cm}
{\sc M.~Beneke}\\[0.5cm]
\vspace*{0.3cm} {\it Randall Laboratory of Physics\\
University of Michigan\\ Ann Arbor, Michigan 48109, U.S.A.}\\[2.4cm]
\vspace*{0.3cm}
 
{\bf Abstract}\\[0.3cm]
\end{center}
The relation between the pole quark mass and the 
$\overline{MS}$-renormalized mass is governed by an infrared 
renormalon singularity, which leads to an ambiguity of order 
$\Lambda_{QCD}$ in the definition of the pole mass. 
We use the renormalization group and heavy quark effective 
theory to determine the exact nature of this singularity up 
to an overall normalization. In the abelian gauge theory, 
the normalization is computed 
partially to next-to-leading order in the 
flavour expansion.\\[1.5cm]

\begin{center} 
\noindent {\it submitted to Physics Letters B}
\end{center} 
 
\newpage
\renewcommand{\thefootnote}{\arabic{footnote}}
\setcounter{footnote}{0}
 

{\bf 1.} It is well known that perturbative series in QCD diverge. 
Recently there has been a resurge of interest in learning about 
nonperturbative effects in a variety of situations through this 
divergence. Among the quantities that have become an object of 
scrutiny is the pole mass of a quark, defined perturbatively as 
the location of the singularity of the renormalized quark propagator. 
The pole mass can be related to the $\MS$ renormalized mass -- 
which in principle can be measured to any accuracy at a very high 
energy scale -- by the series\footnote{A normalization point 
$\mu$ is understood for $m_{\MS}$ and the $\MS$ coupling $\alpha$.}

\be\label{series}
m_{\pole} = m_{\MS} + \sum_{n=0}^\infty r_n \alpha^{n+1}\, 
\ee

\n where the first two coefficients $r_0$ and $r_1$ are known 
\cite{GRA90}. Being infrared (IR) finite and scheme-independent
\cite{TAR81}, the pole mass is the natural mass parameter 
in heavy quark systems, whose scale is governed by the mass of 
the heavy quark. Moreover, the pole mass is implicit in the 
construction of heavy quark effective theory (HQET), which has 
become the major theoretical framework for the description of 
heavy hadron decays \cite{GRI92}. 

Recently, investigation of the set of diagrams symbolized by 
Fig.~1a revealed that in large orders, the coefficients $r_n$ 
are strongly contributed by IR momenta in the integration over 
the gluon line and diverge as \cite{BEN94,BIG94}

\be\label{leading}
r_n \stackrel{n\rightarrow\infty}{=}\frac{C_F}{\pi}\,e^{5/6}\,
\mu\,(-2\beta_0)^n n! ,
\ee

\n where $\beta_0$ is the first coefficient of the $\beta$-function 
[We use $\beta(\alpha)\equiv \mu^2 (\partial \alpha)/(\partial \mu^2) 
= \beta_0\alpha^2 (1+\sum_{n=1} c_n \alpha^n)$. $C_F=4/3$ for SU(3) and 
$C_F=1$ for U(1).]. If one attempts to sum the divergent series 
with the help of the Borel transform, the Borel sum

\be\label{borel}
m_{\pole} = m_{\MS} + \int\limits_0^\infty\mbox{d} t \,e^{-t/\alpha}
\,B[m_{\pole}](t)\qquad B[m_{\pole}](t)\equiv \sum_{n=0}^\infty 
r_n \frac{t^n}{n!} , 
\ee

\n is ambiguous of order $\Lambda_{\rm QCD}$ from a simple 
IR renormalon pole at 
$t=-1/(2\beta_0)$. In the following we adopt the convention that all 
quantities that have divergent expansions are understood as Borel 
sums as in Eq.~(\ref{borel}) with the contour deformed above the 
singularities on the positive axis. Then the ambiguity of order 
$\Lambda_{\rm QCD}$ is equivalently expressed as an imaginary part 
of $m_{\pole}$ of this order.

The ambiguity in the definition of the pole mass can immediately be 
translated into an ambiguity of the parameter $\bar{\Lambda}_{H_Q}$, 
which appears in HQET, if defined as usual through the expansion 
of the heavy hadron mass 

\be \label{lambdabar}
m_{H_Q} = m_{\pole} + \bar{\Lambda}_{H_Q} + O(1/m_{\pole}) .
\ee

\n Physically, this ambiguity of $\bar{\Lambda}_{H_Q}$ arises, because 
the contribution of the spectator to the mass of the hadron can not 
be separated from self-energy corrections to the heavy quark, once these 
become sensitive to the long range part of the Coulomb potential. This 
fact is known in a different language to the lattice community, where 
the extraction of the constant term in the heavy quark potential has faced 
difficulties precisely because of the contamination from self-energy 
contributions. 

In this letter, we address a more formal issue and explore to what 
extent the large order behaviour of the coefficients $r_n$ can be 
determined without any approximation. In general, one expects the 
behaviour of Eq.~(\ref{leading}) to be modified to 

\be\label{general}
r_n \stackrel{n\rightarrow\infty}{=} N\,\mu\,(-2\beta_0)^n \,\Gamma(n+1+b)\left(1+\sum_{k=1}^\infty\frac{s_k}{n^k}\right) ,
\ee

\n turning the simple pole of $B[m_{\pole}](t)$ at $t=-1/(2\beta_0)$ 
into a branch point of strength $1+b$. We shall determine the exact 
nature of the singularity, i.e. $b$ and the $s_k$, entirely in 
terms of the $\beta$-function. The analoguous coefficient $b$ is 
known to date only for the QCD corrections to $e^+ e^-$ annihilation 
and the GLS sum rule \cite{MUE85,ZAK92}, where, however, the corresponding 
singularity is not responsible for the dominant divergent behaviour of 
the series. In these cases one may exploit the 
short distance expansion of the appropriate current product and a 
relation to higher dimension operators. The short distance expansion 
is not available for the pole mass. Instead we will utilize the 
observation of Ref.~\cite{BEN94}, that the IR renormalon at $t=-1/
(2\beta_0)$ in the pole mass is closely related to an {\it ultraviolet} 
(UV) renormalon singularity at the same position in the self-energy of a 
static quark. The normalization $N$ remains elusive. We calculate it 
partially to 
next-to-leading order in the flavour expansion in the abelian theory.

Though we doubt that our results can be used phenomenologically 
to improve the perturbative relation Eq.~(\ref{series}), we believe 
that the arguments that lead to the characterization of 
Eq.~(\ref{general}) are interesting by themselves and may shed 
more light on the simplicities and complexities of divergent series 
in renormalizable theories.\\

{\bf 2.} In this section we consider the self-energy of a static 
quark, described by the leading order effective Lagrangian of 
HQET. Its Lorentz covariant form is 

\be \label{hqet}
{\cal L}_{\rm eff} = \bar{h}_v i v\cdot D h_v + {\cal L}_{\rm 
light} ,
\ee

\n where $v$ is the quark velocity and ${\cal L}_{\rm light}$ the 
QCD Lagrangian for the light degrees of freedom. In contrast to the 
self-energy of a quark of finite mass, the self-energy of a static 
quark is linearly UV divergent. Starting from a bare Lagrangian of 
the above form, renormalization will in general involve a counterterm 
$\bar{h}_v h_v$, such that the renormalized Lagrangian contains a 
residual mass term $\delta m_{\rm res}\propto \mu$ for the static 
quark. We choose a minimal subtraction (MS) like subtraction scheme 
(i.e. $\MS$ or schemes that differ from $\MS$ by a change of scale 
$\mu$). Then a residual mass does not arise to all orders in perturbation theory, but the subtractions are such that the series expansion of 
$\Sigma_{\rm eff}$ has an UV renormalon divergence analogous to 
Eq.~(\ref{general}) \cite{BEN94}. We emphasize the 
choice of MS-like schemes. 
Large order coefficients are scheme-dependent and in general 
(mass-dependent) schemes, this particular UV renormalon divergence can be 
eliminated at the price of introducing a residual mass, which is 
adjusted precisely to this divergence.

At this point we return to the old observation by Parisi \cite{PAR78} 
that the imaginary parts of the Borel-summed Green functions due to 
an UV renormalon at position $t=n/\beta_0$ in the Borel plane are 
proportional to a sum over zero-momentum insertions of all possible 
local operators of dimension $2 n+4$, which can be 
considered as a simple version of a factorization theorem. Generically

\be\label{parisi} 
{\rm Im}\,G(p_k) = \sum_d\sum_{{\rm dim}\,{\cal O}_i=d} 
E_{di}\, G_{{\cal O}_i}\,(p_k) .
\ee

\n Although we are not aware of a general proof of this statement, we 
note that it has been confirmed in a nontrivial case of the photon 
vacuum polarization \cite{VAI94,CEC94} as far as dimension six 
operators are concerned. In fact, Eq.~(\ref{parisi}) is almost intuitive, 
if one understands the emergence of UV renormalons in the context 
of renormalization in general. After the theory is regulated with a 
dimensionful cutoff, renormalization can be expressed as hiding all 
divergent terms in a large-cutoff expansion into a redefinition of the 
low-energy parameters. The remaining cutoff-dependence determines the 
remaining sensitivity to small distances, which is suppressed by a 
power of the cutoff. It is these ``left-overs'', which give rise to 
divergent series, when the series is expressed in terms of the 
renormalized low-energy coupling. However, in principle, order by order 
in the inverse cutoff 
the remaining sensitivity to small distances could be removed by 
adding local higher dimension operators to the Lagrangian, a procedure 
that is indeed widely used in the construction of improved lattice 
actions. By the same token, the imaginary parts created by UV renormalons 
can be described by Eq.~(\ref{parisi}). 

In the particular case of the self-energy, the situation is in fact 
much simpler. The UV renormalon at $t=-1/(2\beta_0)$ is related to 
the dominant (linear) cutoff-dependent term and one does not have 
to deal with a small left-over from renormalization. As indicated by 
the position of the pole and the discussion above, the relevant operators 
have dimension three. There exists only one dimension three operator, 
$\bar{h}_v h_v$, and the general expression Eq.~(\ref{parisi}) reduces 
to

\be\label{insertion}
{\rm Im}\,\Sigma_{\rm eff}\left(vk,\mu,\alpha\right) = E\left(\mu,
\alpha\right)\,\Sigma_{{\rm eff},\bar{h}_v h_v}\!\left(\frac{
vk}{\mu},\alpha\right) ,
\ee

\n as far as the renormalon at $t=-1/(2\beta_0)$ alone is concerned. 
Here $\Sigma_{{\rm eff},\bar{h}_v h_v}$ denotes the self-energy 
with one zero-momentum insertion of $\bar{h}_v h_v$. Eq.~(\ref{insertion}) 
is illustrated to leading order in Fig.~2. 
Comparing the renormalization group equations for $\Sigma_{\rm eff}$ 
(and therefore its imaginary part) and $\Sigma_{{\rm eff},\bar{h}_v h_v}$, 
one finds that the coefficient satisfies 

\be\label{rge}
\left(\mu^2 \frac{\partial}{\partial\mu^2} +  \beta(\alpha)\frac{\partial}{\partial\alpha} - 
\gamma_{\bar{h}_v h_v}(\alpha)\right)\, E\left(\mu,\alpha
\right) = 0 .
\ee

\n The anomalous dimension $\gamma_{\bar{h}_v h_v}(\alpha)$ of 
the operator $\bar{h}_v h_v$ vanishes to all orders in perturbation 
theory, because the operator is related to the conserved current 
$\bar{h}_v \gamma_\mu h_v$ by spin symmetry. A different way to see 
this is to observe that $m_Q \bar{h}_v h_v$ is renormalization group 
invariant, but the HQET expansion parameter $m_Q$ does not run. By 
dimensional arguments, the explicit $\mu$-dependence of $E$ must be 
an overall factor $\mu$ and the solution of Eq.~(\ref{rge}) is 

\be\label{solution}
E\left(\mu,\alpha\right) = {\rm const}\times \mu\,\exp\left(\,
\int\limits_\alpha\mbox{d}\alpha^\prime\frac{1}{2\beta(
\alpha^\prime)}\right) = {\rm const}\times\,\Lambda_{\rm QCD} .
\ee

\n Together with Eq.~(\ref{insertion}) this determines completely 
the UV renormalon singularity of the self-energy up to an overall 
constant in terms of the $\beta$-function and the perturbative expansion 
of $\Sigma_{{\rm eff},\bar{h}_v h_v}$. Note that it depends on 
$vk/\mu$ only through $1/n$-corrections to the leading asymptotic 
behaviour in accordance with the expectation that the divergence of 
the derivative of the self-energy is suppressed, because it is linearly 
divergent only through linearly divergent subdiagrams. For later use, 
we note the equality

\be\label{equality}
\Sigma_{{\rm eff},\bar{h}_v h_v} = 1 - \frac{\partial \Sigma_{\rm eff}}
{\partial vk} ,
\ee

\n which can easily be shown diagrammatically by routing the external 
momentum $k$ through the heavy quark line. The unity on the r.h.s. 
appears, because the tree diagram on the r.h.s. of Fig.~2 is included 
in the definition of $\Sigma_{{\rm eff},\bar{h}_v h_v}$.\\

{\bf 3.} We now relate the UV renormalon in the self-energy of a 
static quark discussed above to the leading IR renormalon in the 
pole mass, generalizing the leading order observation of 
Ref.~\cite{BEN94}. To this end we sandwich the full inverse 
propagator $S^{-1}$ of a quark in QCD between two projectors 
$P_v=(1+\!\!\not\! v)/2$ and define

\be P_v \,S^{-1}\!\left(p,m_{\MS},\mu;\alpha\right)\,P_v \equiv 
P_v\,S_{\rm P}^{-1}\!\left(v k,m_Q,\mu;\alpha\right)
\ee

\n with $p=m_Q v+k$. As long as $p$ is real and away from the 
zero of $S^{-1}$, (the Borel sum of) $S^{-1}$ does not have an 
imaginary part of order $\Lambda_{\rm QCD}$, 
because its Borel transform has no singularity at 
$t=-1/(2\beta_0)$. Odd powers of 
$\Lambda_{\rm QCD}$ can arise only from on-shell Feynman integrals 
or expansion around mass-shell. For the purpose of our argument, 
we shall assume the hierarchy $m_Q\gg vk\gg\Lambda_{\rm QCD}$. Then 
the largest imaginary part of $S_{\rm P}^{-1}$ comes from a 
singularity of its Borel transform at $t=-1/\beta_0$, which is 
or order $\Lambda_{\rm QCD}^2/v k$, parametrically smaller than 
$\Lambda_{\rm QCD}$. 

We wish to expand $S_{\rm P}^{-1}$ in $vk/m_Q$ about $v k=0$, 
where $p$, $m_Q$ and $v k$ should be real. Anticipating an 
imaginary part of the Borel sum of $m_{\rm pole}$ of order 
$\Lambda_{\rm QCD}$, we define 

\be m_Q\equiv m_{\pole}-\delta \hat{m}\qquad 
\delta\hat{m}\equiv i \delta m\equiv i \,{\rm Im}\,m_{\pole} .
\ee

\n To all orders in perturbation theory, $S_{\rm P}^{-1}$ can be 
matched onto the self-energy of a static quark computed from 
HQET without a residual mass term, Eq.~(\ref{hqet}), and 
expansion parameter $m_Q=m_{\rm pole}$. 
Thus defining $k^\prime=k-\delta\hat{m} v$ and 
using $\delta m\ll v k$, one obtains

\bea\label{matching}
S_{\rm P}^{-1}\left(v k,m_Q,\mu\right)\!\!&=&\!\!
C\left(\frac{m_{\rm pole}}{\mu}\right) \left(v k^\prime - 
\Sigma_{\rm eff}\left(v k^\prime,\mu\right)\right) + O\left(
\frac{|k^\prime|^2}{m_{\rm pole}}\right)\nonumber\\
\hspace*{-1cm} &=&\!\! 
C\left(\frac{m_Q}{\mu}\right) \left[v k - 
\Sigma_{\rm eff}\left(v k,\mu\right) - \delta\hat{m} \left(
1 - \frac{\partial \Sigma_{\rm eff}}{\partial vk}(vk/\mu)\right) 
\right]\\
\hspace*{-1cm} &&\!\!+\, 
\frac{\delta\hat{m}}{\mu}\,C^\prime\left(\frac{m_Q}{\mu}\right) 
\left[v k - \Sigma_{\rm eff}\left(v k,\mu\right)\right]
+ O\left(\frac{(\delta m)^2}{m_Q},
\frac{|k|^2}{m_Q}\right) ,\nonumber 
\eea

\n with a matching coefficient $C$ that contains all $m_Q$-dependence. 
While $\Sigma_{\rm eff}$ contains an imaginary part of order 
$\Lambda_{\rm QCD}$, the l.h.s. of Eq.~(\ref{matching}) does not. 
Investigating the scaling of all potential imaginary parts in 
quantities on the r.h.s. of Eq.~(\ref{matching}) with $vk$, $\mu$ and 
$\Lambda_{\rm QCD}$, one finds that the condition that imaginary 
parts of order $\Lambda_{\rm QCD}$ cancel, can be expressed as

\be {\rm Im}\,\left[v k - 
\Sigma_{\rm eff}\left(v k,\mu\right) - \delta\hat{m} \left(
1 - \frac{\partial \Sigma_{\rm eff}}{\partial vk}(vk/\mu)\right) 
\right] = 0 .
\ee

\n Combining this condition with Eqs.~(\ref{insertion}) and 
(\ref{equality}), we obtain the final result

\bea \label{final}
 {\rm Im}\,m_{\rm pole}\!\!&=&\!\! \delta m = - E\left(\mu,\alpha\right) 
= {\rm const}\times\,\Lambda_{\rm QCD}\\
&=&\!\!{\rm const}\times\,\mu\,e^{1/(2\beta_0\alpha)} \,
\alpha^{c_1/(2\beta_0)}\left(1 + \frac{c_2-c_1^2}{2\beta_0} 
\alpha + \ldots
\right)\nonumber, 
\eea

\n where $E$ is defined by Eq.~(\ref{insertion}) and given by 
Eq.~(\ref{solution}). The second line of Eq.~(\ref{final}) 
is easily translated into the form of the singularity of 
$B[m_{\rm pole}](t)$ at $t=-1/(2\beta_0)$, and therefore into the  
large order behaviour of the coefficients $r_n$, Eq.~({\ref{general}), 
which relates the pole mass to the $\MS$ mass\footnote{For this 
argument, we assume that no ``unknown'' singularities are present. 
Then the IR renormalon at $t=-1/(2\beta_0)$ lies closest to the 
origin of the Borel plane and determines the asymptotic behaviour 
of the perturbative coefficients.}. From Eq.~(\ref{solution}) or 
(\ref{final}) one deduces that except for the overall normalization 
$N$ all quantities that characterize Eq.~(\ref{general}) are 
determined by the $\beta$-function. For instance,

\be\label{eg} 
b=-\frac{c_1}{2\beta_0}\qquad s_1 = \frac{c_1^2-c_2}{4\beta_0^2} ,
\ee

\n and, in general, $1/n^k$-corrections to the leading asymptotic 
behaviour involve the $\beta$-function coefficients up to $c_{k+1}$. 
We stress that these conclusions follow with almost no dynamical 
input. We used essentially only the linear UV divergence of the 
self-energy of the static quark.  The only ``nonperturbative'' parameter 
that remains and is not fixed by very general arguments, is the overall 
normalization \cite{GRU93,BEN93a}. This is because a nonperturbative 
approach comparable to the semiclassical expansion for instanton-induced 
(combinatorically) divergent series is lacking for renormalons.

For the ambiguity present from the perturbative definition of the 
pole mass, Eq.~(\ref{final}) shows that it is proportional to $\Lambda_{
\rm QCD}$ with no multiplying logarithm and series in $\ln(\mu/\Lambda_{
\rm QCD})$, which is in fact the only possible form compatible with 
scheme-invariance of the pole mass. In our formal derivation, the 
absence of logarithms is due to the vanishing anomalous dimension of 
$\bar{h}_v h_v$, which indeed is directly related to the renormalization 
group invariance of the pole mass as noted above. The bonus of the 
formal derivation is Eq.~(\ref{insertion}), which allows to calculate 
(approximately) the overall normalization from the self-energy in HQET.

As mentioned before, the renormalon ambiguity in $m_{\pole}$ translates 
into an ambiguity in the $\bar{\Lambda}$-parameter of HQET, see 
Eq.~(\ref{lambdabar}). It is tempting to estimate this ambiguity 
from a leading order calculation of the normalization. The arbitrariness 
of this procedure is reflected in the different values, spanning 
$\delta m_{\pole} = \delta\bar{\Lambda}_{H_Q} = (50 - 300)\,$ MeV, 
quoted in Refs.~\cite{BEN94,BIG94} and the present considerations 
can not improve the expectation that this ambiguity is of order 
$\Lambda_{\rm QCD}$.\\

{\bf 4.} In a nonabelian theory, a systematic approach to calculate the 
normalization of renormalons is not known. Explicit calculations can be 
performed in an abelian toy model (QED with $N_f$ massless fermion for 
${\cal L}_{\rm light}$) in an expansion in the number of flavours\footnote{
The same expansion can also be applied to the nonabelian theory, but 
can not be taken seriously numerically, since it is singular at 
$N_f=33/2$, when $\beta_0$ is zero.}. To organize this expansion, one 
defines $a=N_f\alpha$ and expands in $a$ and $1/N_f$. The $\beta$-function 
for the rescaled coupling is $\beta(a)=b_0 a^2 (1+\hat{c}_1 a + \ldots)$, 
$b_0=1/(3\pi)$ and $\hat{c}_1=3/(4\pi)$. We also expand the normalization 

\be N = N_0 \left(1 + \frac{N_1}{N_f} + \ldots\right)\qquad
N_0 = \frac{C_F}{\pi N_f}\,e^{5/6} ,
\ee

\n where $N_0$ is minus the asymptotic behaviour of coefficients 
obtained from the set of self-energy diagrams in Fig.~1a. To next-to-leading 
order in $1/N_f$, one has to extract the asymptotic behaviour from 
the diagrams symbolized by Fig.~1b-d, which according to 
Eqs.~(\ref{general}) and (\ref{eg}) is of the form

\be\label{nlo}
r_n\stackrel{n\rightarrow\infty}{=} N\,\mu\,(-2\beta_0)^n 
\left[1 + \frac{1}{N_f}\left\{-\frac{\hat{c}_1}{2\beta_0}\ln n + 
N_1\right\} + O\left(\frac{1}{N_f^2}\right)\right] \times \left(
1 + O\left(\frac{1}{n}\right)\right) .
\ee

\n A simplification arises, because diagrams 1c and 1d do not 
contribute to the overall normalization, when only one 
of the two gauge boson momenta is large. For example, diagram 1c
amounts to a radiative correction of the r.h.s. of Fig.~2 and therefore 
affects the asymptotic behaviour of $\Sigma_{eff}$ only at the 
level of $1/n$-corrections, whereas diagram 1d does not contribute to 
the UV renormalon at $t=-1/(2\beta_0)$ at all, because it has no 
linearly divergent subdiagrams. The present case is markedly distinct 
from -- and simpler than -- the 
vacuum polarization in QED and its first UV renormalon at 
$t=1/\beta_0$, where diagrams with two chains of fermion 
bubbles and one large internal momentum are not suppressed \cite{VAI94}. 
The reason is that for the self-energy, we are interested in the dominant large-momentum contribution to the Feynman integrals. A contribution 
to $N_1$ from diagrams 1c and d can only come from the region, where 
both gauge boson momenta are large and of the same order. 

Diagrams 1a and b are given by the first two terms in the 
$1/N_f$-expansion of diagram 1a with integration 
over the full gauge boson propagator.
The Borel transform (defined analogous 
to Eq.~(\ref{borel})) of the class of diagrams obtained by insertion of 
the full propagator is

\be\label{full}
B[\Sigma_{\rm eff}^{\rm a+b}](t) = - \frac{4\pi C_F}{i N_f} 
\int\frac{\mbox{d}^4 q}{(2\pi)^4} \,\frac{(vk)^2-q^2}{v (k+q)\, q^4}
\,B\left[\frac{\alpha}{1+\Pi(-q^2/\mu^2,\alpha)}\right](t) ,
\ee

\n as discussed in a general context in Ref.~\cite{GRU93}. Expanding 
the gauge vacuum polarization $\Pi$, 

\be \Pi(q^2) = \Pi_0(q^2) + \frac{1}{N_f}\,\Pi_1(q^2) + \ldots
\ee
\be
\Pi_0(q^2) = - b_0\alpha \ln X\qquad X\equiv -\frac{q^2}{\mu^2} 
e^{-5/3} ,
\ee

\n one finds

\be 
B\left[\frac{\alpha}{1+\Pi(q^2)}\right]\!(t) = X^{b_0 t} - \frac{1}{N_f} 
\int\limits_0^t\mbox{d} s\,s X^{b_0 s} B\left[\frac{\Pi_1}{\alpha}
\right]\!(t-s) + O\left(\frac{1}{N_f^2}\right) ,
\ee

\n where the first term corresponds to diagram 1a and the second to 
1b. The Borel transform of $\Pi_1/\alpha$ is known \cite{BEN93b} and 
can be written as 

\be B\left[\frac{\Pi_1}{\alpha}\right]\!(t) = X^{b_0 t} F(t) - G(t) ,
\ee

\n where $F(t)$ is a scheme-independent function given by 
\cite{BEN93b,BRO93,BEN94}

\be\label{f}
F(t) = \frac{8}{3\pi^2}\frac{1}{1-(1+b_0 t)^2} 
\sum_{k=2}^\infty\frac{(-1)^k k}{(k^2-(1+b_0 t)^2)^2}
\stackrel{t\rightarrow 0}{=} -\frac{\hat{c}_1}{t} .
\ee

\n The divergence of $F(t)$ at $t=0$ is regulated by $G(t)$, which is 
scheme-dependent and without poles in $t$. 
But, since $\alpha/(1+\Pi(q^2))$ is scheme-independent 
in QED, the scheme-dependence of its Borel transform is compensated 
by the scheme-dependence of the coupling in the exponent of the 
inverse Borel transform. Consequently, the subtraction dependence 
of $G(t)$ can be expressed in terms of the $\beta$-function. Define 
$F_{\rm reg}(t)\equiv F(t) + \hat{c}_1/t$, 
$G_{\rm reg}(t)\equiv G(t) + \hat{c}_1/t$, such that both $F_{\rm reg}(t)$ 
and $G_{\rm reg}(t)$ are finite at $t=0$ and 

\bea 
I(u) \!\!&\equiv&\!\! - \frac{4\pi C_F}{i N_f} 
\int\frac{\mbox{d}^4 q}{(2\pi)^4} \,\frac{(vk)^2-q^2}{v (k+q)\, q^4}
\,X^{-u}\\
&=&\!\! \frac{4\pi C_F}{N_f}\,vk \left(-\frac{2 vk}{\mu}\right)^{-2 u} 
e^{5 u/3}\,(-6)\,\frac{\Gamma(-1+2 u)\Gamma(1-u)}{\Gamma(2+u)} .
\nonumber\eea

\n Then 

\bea
B[\Sigma_{\rm eff}^{\rm a+b}](t)\!\!&=&\!\! I(-b_0 t)\left[1-
\frac{1}{N_f}\int\limits_0^t\mbox{d} s\,s F_{\rm reg}(t-s)\right]
+ \frac{1}{N_f}\int\limits_0^t\mbox{d} s\,s I(-b_0 s)
G_{\rm reg}(t-s)\nonumber\\
&&\!\! +\,\frac{\hat{c}_1}{N_f}\int\limits_0^t\mbox{d} s\,
\frac{s}{t-s} \left(I(-b_0 t)-I(-b_0 s)\right) .
\eea

\n It is simple to extract the leading singular behaviour as 
$t\rightarrow -1/(2 b_0)$ from this equation. The term 
involving $G_{\rm reg}$ produces a $\ln \,(1+2 b_0 t)$ singularity 
and contributes only to $1/n$-corrections in Eq.~(\ref{nlo}), 
consistent with the scheme-independence of $N_1$. The result 
is 

\be\label{singularity} 
B[\Sigma_{\rm eff}^{\rm a+b}](t) \stackrel{t\rightarrow 
-1/(2 b_0)}{=} -\frac{C_F}{\pi N_f}\,e^{5/6}\,\frac{\mu}
{1+2 b_0 t} \left[1+\frac{1}{N_f}\left\{ \frac{\hat{c}_1}
{2 b_0}\,\ln\,(1+2 b_0 t) + A\right\} + \ldots \right]
\ee

\n with

\bea A\!\!&\equiv&\!\! \frac{\hat{c}_1}{2 b_0} - 
\!\!\!\!\int\limits_0^{-1/(2 b_0)}
\!\!\!\!\mbox{d} s\,s F_{\rm reg} \,(-1/(2 b_0)-s)
= \frac{27}{8}\ln\frac{4}{3} \\
&&\!\! + 12\sum_{k=2}^\infty 
\frac{(-1)^k k}{(k^2-1)^2} \Bigg[\frac{1}{2 k^2} - 
\ln\frac{k^2-1}{k^2-1/4} + \frac{1-3 k^2}{2 k^3} 
\left({\rm artanh}\, k^{-1}-{\rm artanh\,} (2 k)^{-1}\right)
\Bigg] \nonumber
\eea

\n The last step consists in converting the singularity of the 
Borel transform in Eq.~(\ref{singularity}) into the asymptotic 
behaviour of the coefficients $r_n$. We reproduce correctly the 
coefficient of the $\ln n$-term in Eq.~(\ref{nlo}), which can 
already be deduced from Eq.~(\ref{final}) without any calculation,  
and find ($\gamma_E$ is Euler-Mascheroni's constant)

\be N_1^{\rm b} = A - \frac{\hat{c}_1}{2 b_0} \gamma_E = 0.6831...
\ee

\n This concludes the 
partial evaluation of the normalization in next-to-leading 
order in the abelian theory.  Note that to obtain $N_1$ one needs 
the exact vacuum polarization $\Pi_1$, in accordance with general 
expectations \cite{GRU93}. We repeat that a potential contribution 
from diagram c and d from the region where both internal momenta 
are large is still missing. The presented calculation of 
$N_1$ should be considered as an illustration how the most complicated 
diagram b can be easily evaluated, using known results on the 
vacuum polarization. The Borel transform of diagrams c and d 
is given by ($u=-b_0 t$, $v=-b_0 s$)

\bea
B[\Sigma_{\rm eff}^{\rm c+d}](t)\!\!&=&\!\!\left(\frac{3}{2\pi N_f}
\right)^2 e^{-5 u/3}\,v k \left(-\frac{2 v k}{\mu}\right)^{-2 u}
\int\limits_0^t\mbox{d} s \,\frac{\Gamma(-1+2 v)\Gamma(1-v)
\Gamma(1-u+v)}{\Gamma(2+v)\Gamma(2+u-v)}\nonumber\\
&&\!\!\times \left[\frac{\Gamma(-1+2 u)}{\Gamma(1+2 v)} 
+2\,\frac{\Gamma(-2+2 u)}{\Gamma(2 v)} - \Gamma(-1+2 u-2 v)\right] .
\eea

\n However, before one can use this expression, one has to extend 
the renormalization of the Borel transform \cite{BEN94} to diagrams 
with two chains.\\

{\bf 5.} The pole mass or, respectively, the self-energy of a 
static quark is probably the simplest quantity, for which the leading 
renormalon can be determined up to a global normalization. Its 
simplicity might also open a path to tackle the outstanding problem 
to identify which diagrams in addition to fermion bubbles are 
responsible for converting the abelian $\beta_0$ to its nonabelian 
value, which is expected on general grounds in Eq.~(\ref{general}). 
To some extent, this issue can be addressed at order $1/N_f^2$ in 
the flavour expansion of the nonabelian theory. The obvious replacement 
of one fermion bubble in the diagram 1a by a gluon or ghost bubble 
is not sufficient to produce the $1/N_f^2$-contribution from expanding 
$\beta_0^n$. It would be interesting to check that the two-loop 
diagram involving the three-gluon vertex dressed by fermion bubbles 
can make up for the missing piece.\\

{\bf Acknowledgements.} I wish to thank V.~M.~Braun 
for discussions and acknowledge support by the 
Alexander-von-Humboldt Foundation.
 
 

 
\end{document}